\documentclass[a4paper]{article}

\usepackage{url}
\usepackage{epsfig}
\usepackage{subfigure}
\usepackage{calc}
\usepackage{amssymb}
\usepackage{amstext}
\usepackage{amsmath}
\usepackage{amsthm}
\usepackage{pslatex}
\usepackage[small]{caption}

\usepackage{graphicx}
\usepackage{amssymb}

\usepackage{tikz}
 	\usetikzlibrary{shapes.misc,intersections}
\tikzset{idnode/.style={inner ysep = 4pt,inner xsep = 8pt, draw}}
\tikzset{varnode/.style={idnode,rounded rectangle}}
\tikzset{detnode/.style={idnode,rounded rectangle,double}}
\tikzset{utilnode/.style={idnode, chamfered rectangle, chamfered rectangle xsep=32pt, chamfered rectangle ysep=0pt}}
\tikzset{actionnode/.style={idnode,rectangle}}
\tikzset{cond/.style={>=stealth,draw}}
\usepackage{pgfplots}
\usepackage{acronym}
\acrodef{PGM}{Probabilistic Graphical Model}
\acrodef{ML}{Machine Learning}
\acrodef{BDI}{Beliefs, Desires and Intentions}
\acrodef{SRL}{Statistical Relational Learning}
\acrodef{PLP}{Probabilistic Logic Programming}
\acrodef{ASL}[\textsc{ASL}]{\textsc{AgentSpeak(L)}}
\acrodef{BN}{Bayesian Network}
\acrodef{AI}{Artificial Intelligence}
\acrodef{PCA}{Principal Component Analysis}
\acrodef{DNN}{Deep Neural Networks}
\acrodef{CPD}{Conditional Probability Distribution}
\acrodef{MEU}{Maximum Expected Utility}
\newcommand{\JASON}{\textsc{Jason}}
\newcommand{\GM}{\textsc{GoldMiners}}
\newcommand{\WEKA}{\textsc{weka}}
\newcommand{\SAMIAM}{\textsc{samiam}}
\newcommand{\PROLOG}{\textsc{ProLog}}

\begin{document}
\title{Probabilistic Selection in \textsc{AgentSpeak(L)}}
%\subtitle{the layout of a research project}}
%

%
\author{%
Francisco~Coelho\\
\texttt{fc@di.uevora.pt}
\and%
Vitor~Nogueira\\
\texttt{vbn@di.uevora.pt}
}
\date{}
%
%
% \affiliation{
% \sup{1} Departamento de Inform\'{a}tica, Universidade de \'{E}vora, Rua Rom\~{a}o Ramalho 58, 7000-671 \'{E}vora, Portugal
% }
% %
% \affiliation{
% \sup{2}Laboratory of Agent Modelling (LabMAg)
% }
% %
% \email{\{fc, vbn\}@di.uevora.pt}
% }
%\onecolumn
\maketitle
%\normalsize
%\vfill

%

%
\begin{abstract}
Agent programming is mostly a symbolic discipline and, as such, draws little benefits from probabilistic areas as machine learning and graphical models. However, the greatest objective of agent research is the achievement of autonomy in dynamical and complex environments --- a goal that implies embracing uncertainty and therefore the entailed representations, algorithms and techniques.
This paper proposes an innovative and conflict free two layer approach to agent programming that uses already established methods and tools from both symbolic and probabilistic artificial intelligence.  Moreover, this framework is illustrated by means of a widely used agent programming example, \GM.
\end{abstract}
%

%
%\keywords{Agent Programming, Symbolic AI, Probabilistic AI}
%
\section{\uppercase{Introduction}}
\noindent
Agent autonomy is a key objective in \ac{AI}. Complex and dynamic environments, like the physical world where robots must delve, impose a degree of uncertainty that challenges the vocation of symbolic processing. But while a probabilistic approach --- currently expressed in \ac{ML} and \acp{PGM} \cite{koller2009probabilistic}--- is required for certain aspects of such tasks, a great deal of agent programming is better handled by declarative programming (\emph{e.g.} \PROLOG) and more specifically, \ac{BDI} architectures for autonomous agents, part of symbolic \ac{AI}.%\AUTHNOTE{\textbf{language}~ are \emph{symbolic}, \emph{logic} and \emph{declarative} being properly used?}

\section{\uppercase{State of the art}}

\begin{figure*}[t]
\begin{center}
\begin{tikzpicture}
\node at (-1,2) (ENV1) {Percepts};
\node[draw] at (0.5,2) (BRF) {BRF};
\node[draw, rounded corners] at (2.5,3) (BB) {Beliefs};
\node[draw, rounded corners] at (2.5,1) (EV) {Events};
\node[draw, rounded corners] at (2.5,2) (PL) {Plans library};
\node[draw,fill=gray!50!white] at (5,1) (SE) {Select event};
\node[draw] at (5,3) (GO) {Generate options};
\node[draw,fill=gray!50!white] at (8,3) (SO) {Select option};
\node[draw, rounded corners] at (8,2) (IQ) {Intentions};
\node[draw,fill=gray!50!white] at (8,1) (SI) {Select intention};
\node at (10,2) (ENV2) {Action};
\path[->, every edge/.style={cond}]
	(ENV1) edge (BRF)
%	(BRF) edge (BB)
%	(BB) edge (BRF)
%	(BRF) edge (EV)
	(EV) edge (SE)
%	(PL) edge (GO)
	(SE) edge (GO)
	(BB) edge (GO)
	(GO) edge (SO)
	(SO) edge (IQ)
	(IQ) edge (SI)
%	(SI) edge (ENV2)
;
\draw[cond,->,rounded corners=3pt] (PL) -| (GO);
\draw[cond,->,rounded corners=3pt] (BRF) |- (EV);
\draw[cond,<->,rounded corners=3pt] (BRF) |- (BB);
\draw[cond,->,rounded corners=3pt] (SI) -| (ENV2);
\end{tikzpicture}
\end{center}
\caption{The \JASON\ deliberation process very resumed, with selection functions highlighted. The \ac{ASL} only specifies the signature of the functions omitting any conditions, besides the type, on the output.}
\label{fig:jason.deliberation}
\end{figure*}
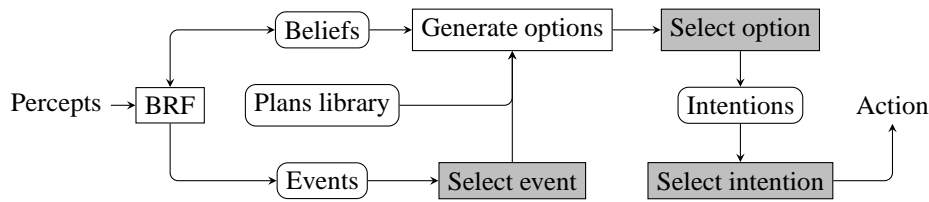{}

Although the symbolic and probabilistic areas of AI correspond, in fact, to different and often antagonist cultures and perspectives, they are not necessarily incompatible.
Bridges between them are being built based on distribution semantics \cite{sato1995statistical} or markov random fields \cite{domingos2006unifying}. From that common ground there are two possible paths towards the interplay of symbolic and probabilistic \ac{AI}: the extension of \acp{PGM} with lo\-gi\-cal and relational  representations (done by  \ac{SRL}) \cite{sato2011general,sato2007inside} and the extension of logic programming languages with probability, in \ac{PLP} \cite{Fierens:2013fk,fierens2012inference,gutmann2011learning}.

From the point of view of programming autonomous agents the sym\-bo\-lic/pro\-ba\-bi\-lis\-tic division essentially still persists: symbolic architectures, such as \ac{BDI}, describe the behaviour of the agents on the basis of metaphors (\emph{e.g.} goals, beliefs) drawn from human behaviour while the principle of \ac{MEU} is included, as influence diagrams, in probabilistic  \ac{AI} but there is only seminal work blurring that division.

Concerning agents programming \JASON\ \cite{bordini2007programming} is a popular \acf{ASL} \cite{rao1996agentspeak} interpreter and framework, triggering a considerable amount of research (\emph{e.g.} \cite{bordini2010semantics,bordini2006bdi}). The \ac{BDI} architecture in general, including \ac{ASL} and \JASON\ in particular, outline a set of symbolic data structures and processes with more or less detailed semantics. The \ac{ASL} as implemented in \JASON\ specifies that the deliberation cycle, depicted in figure \ref{fig:jason.deliberation}, certain selection steps are handled by certain functions. It also defines the signatures of these functions but omits their inner workings. Such omissions play a central role in this work. Despite some work concerning intention selection \cite{bordini2002agentspeak} the default selection function implementation in \JASON\ is a simple process based in round-robin scheduling: intentions form a stack and at each time-step the head action of the top intention is selected; that intention is then sent to the bottom of the stack. This somewhat simplistic approach to selection is good enough for many tasks, including winning planning competitions \cite{bordini2007using,hubner2008developing}.

\begin{figure}[t]\begin{center}
\begin{tikzpicture}
			\begin{axis}[smooth,
				width = 0.75\textwidth,
				height = 0.46\textwidth,
				title = {Effect of Noise in Performance},
				xlabel = {Sensor noise}, ylabel = {Gathered golds},
				xmin = -0.005, xmax = 0.105,
				xtick = {0.0, 0.025, 0.05, 0.075, 0.1},
				xticklabel style = {
					/pgf/number format/precision = 3,
					/pgf/number format/fixed,
					/pgf/number format/fixed zerofill,
				},
				mark options = {scale = 2,},
				]
				%%%%%%%%%%%%%%%%%%%%%%%%%%%%%%%%%%%%
				%
				%	MEAN PLOTS
				%
				%	dummy
				%
				\addplot[solid,black,mark=diamond]
					table[x=NOISE,y=DUMMY] {experiment2.dat};
				\addlegendentry{dummy};
				%
				%	noc
				%
				\addplot[solid,black,mark=square]
					table[x=NOISE,y=NOC] {experiment2.dat};
				\addlegendentry{smart};
				%
				%%%%%%%%%%%%%%%%%%%%%%%%%%%%%%%%%%%%
				%
				%	MEAN +/- STDVAR PLOTS
				%
				%
				%	dummy
				%
				\addplot[black!10, name path = DMIN]
					table[x=NOISE,y=DUMMYMIN] {experiment2.dat};	
				\addplot[black!10, name path = DMAX]
					table[x=NOISE,y=DUMMYMAX] {experiment2.dat};
				%
				%	noc
				%
				\addplot[black!10, name path = NMIN]
					table[x=NOISE,y=NOCMIN] {experiment2.dat};
				\addplot[black!10, name path = NMAX]
					table[x=NOISE,y=NOCMAX] {experiment2.dat};
				%
				%%%%%%%%%%%%%%%%%%%%%%%%%%%%%%%%%%%%
				%
				%	FILL PLOTS
				%
				%
				%	dummy
				%
				% \addplot[black!10, opacity=0.25]
				% 	fill between [of= DMIN and DMAX];
				% %
				% %	noc
				% %
				% \addplot[black!10, opacity=0.2 	5]
				% 	fill between [of= NMIN and NMAX];
			\end{axis}
		\end{tikzpicture}
\caption{If sensors report misreadings of the environment state the symbolic inference process inherent to \ac{BDI} uses false perceptions as (true) facts of the environment and the deliberation process works on wrong assumptions. This plot relates sensor noise (the rate of sensor misreadings, in the horizontal axis) with agent performance measured by the the number of gathered golds (in the vertical axis). Two teams are plotted, the basic reference ``dummy'' that barely uses \ac{BDI} features and the ``smart'' team, fully \ac{BDI}, (designed by \cite{hubner2008developing}) that won the 2006 ``Multi-agent Programming Contest'' \cite{dastani2007second} featuring the \GM\ scenario. 
Each data point summarizes the number of gathered golds by team in a given noise parameter and consists of the mean and standard variation of ten samples. The mean is traced by a thin black line and standard variation by a band centred in the mean value. Values between data-points are interpolations.
}

\label{fig:experiment1.results}
\end{center}\end{figure}
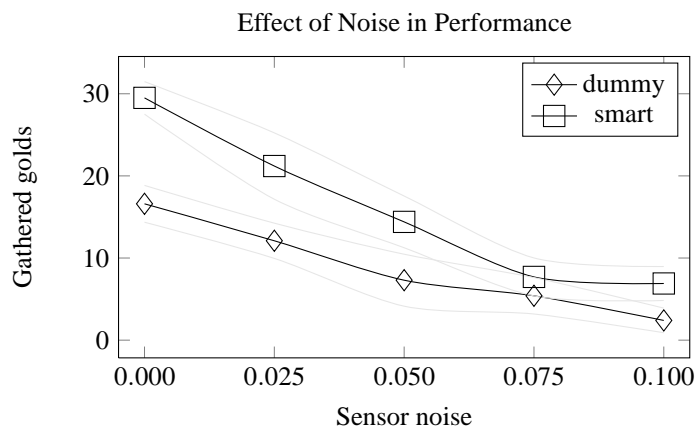
However we can see \JASON\ agents in trouble when their environment  becomes stochastic. This assertion is hinted by a simple experiment plotted in figure \ref{fig:experiment1.results}: the \GM\ is a virtual scenario used in the 2006 Multi-Agent Programming Contest \cite{behrens2011special} edition, now part of \JASON's examples. If this scenario is run unchanged the two playing teams reach scores that are clearly reduced if even a small amount of noise is added ($5\%$ in the plotted experiment) to the perceptions and actions.

It turns out that a \ac{BN} is a natural representation of the complex  interdependency of random variables and, therefore, a great candidate to represent   probabilistic beliefs. But the task of replacing symbolic beliefs by \acp{BN} is far from trivial in part because changing the data structure of beliefs entails a chain of reconsiderations about every aspect of the architecture. For example, plans have contexts that must be unifiable with the agent's beliefs base; changing the beliefs base from a set of closed formulas to a joint distribution of random variables will break (unchanged) unification with those contexts. 

\section{\uppercase{Probabilistic Selections}}

\noindent Currently the problem of extending \ac{ASL} data structures with probabilistic features is being addressed by different authors \cite{luzalternatives,fagundes2009deliberation,fagundes2007integrating,kieling2011insertion,silva2011agentspeak} but isn't yet fully solved.
An alternative and less intrusive application of probabilistic \ac{AI} to \ac{ASL} targets the processes instead of the data. Bounding probabilistic techniques to the computation of \ac{ASL} selection functions (events, options and intentions\footnote{Recent versions of \JASON\ include an inbox, an outbox and a select message function.}, as in figure \ref{fig:jason.deliberation}) promises a number of benefits:
\begin{itemize}
\item selection functions usually have natural formulations in terms of optimization problems which, in many cases, are well handled by probabilistic algorithms;
\item since their computations are unspecified (in \ac{ASL} or \JASON) probabilistic techniques can be used without compromising previous work;
\item symbolic and probabilistic \ac{AI} roles are clearly separated but both simultaneously contribute to the agent behaviour: 
	\begin{itemize}
	\item symbolic programming uses unchanged \ac{ASL} to define high level agent behaviour, with plans, goals, beliefs, resolution,\emph{etc};
	\item probabilistic algorithms use unchanged tools (\emph{e.g.} \WEKA\   \cite{bouckaert2010weka,witten2005data,dimov2007weka,hall2009weka} or \SAMIAM \footnote{From \url{http://reasoning.cs.ucla.edu/samiam/}.})  to process low level noisy signals --- with \acp{BN}, influence diagrams, monte-carlo markov chains, expectation-maximization, \emph{etc};
	\end{itemize}
%\item theoretical results of probabilistic \ac{AI} (\emph{e.g.} concerning error estimation) can be used to better 
\end{itemize}
In summary, defining the selection functions of \ac{ASL} as optimization tasks to be solved by probabilistic techniques seems to pose a promising set of open problems, with the potential of great contribute for an evolution (and not a revolution) of agent programming.
%

%%
%\item What \acp{PGM} techniques apply?
%%
%\item In what tools?
%	\begin{itemize}
%	\item \WEKA~ for machine learning and data analysis
%	\item \SAMIAM~ for graphical models
%	\item others, preferably \texttt{java} based, for better integration with \texttt{Jason};
%	\end{itemize}
%%
%\item How to measure and compare with symbolic only agents?
%	\begin{quote}
%	With the competition \textbf{score}, of course! (yes, again)
%	\end{quote}
%\end{itemize}

\subsection{\uppercase{Problem Statement}}
\noindent
Our immediate goal is to use the original \ac{ASL} \JASON\ programs of the miners team, the stochastic environment of the experiment depicted in figure \ref{fig:experiment1.results} and define the computation of the intention selection with a probabilistic process based in the \ac{MEU} principle. Success achieving this goal can be directly measured by the effect on the team's performance: the greater the performance increase, the greater the success.
This paper motivates and outlines the design of such probabilistic process. Implementation and evaluation are postponed to future developments. 
Next a more precise statement concerning intention selection in \JASON\ is preceded by short overviews of the \GM\ competition, noise model used, miners coordination and influence diagrams.
The original environment of the \GM\ competition is partially observable, stochastic, sequential, dynamic and discrete (see \cite{russell2003artificial} about this classification). The competition description states that ``[t]he environment of the multi-agent system was a grid-like world where agents could move from one cell to a neighbouring cell if there was no agent or obstacle already in that cell. In this environment, gold could appear in the cells. Participating agent teams were expected to explore the environment, avoid obstacles and compete with another agent team for the gold. The agents of each team could coordinate their actions in order to collect as much gold as they could and to deliver it to the depot where the gold can be safely stored. Agents had only a local view on their environment, their perceptions could be incomplete, and their actions could fail [\emph{in}  \cite{dastani2007second}].''
Although noise is present in the original competition scenario in the form of incomplete perception and action failure the \GM\ examples in the \JASON\ distribution (besides providing a proxy to that competition simulator) optionally use a local simulator for development and evaluation purposes. In thar local simulator noise only increases the probability of action failure in proportion to current cargo. Also the environment type changes from stochastic (where gold can appear in cells) to strategic (because the only changes in the environment state are produced by agents).
Experimental results depicted in figure \ref{fig:experiment1.results} result from this local simulator with added noise in the perceptions (number of gold pieces, \emph{etc.}) and action choice. The amount of noise is configurable by an external parameter that when zero the simulation becomes the original and local, noise free, one.
The \GM\ miners team follows a coordination protocol concerning the collection and transport of newly found gold pieces. There are two kinds of agents, a single leader and a set of miners. 
``[The] leader helps the miners to coordinate themselves in [\ldots] the negotiation process that is started when a miner sees a piece of gold and is not able to collect it (because its container is full). This miner broadcasts the gold location to other miners who then send bids to the leader. The leader chooses the best offer and allocate the corresponding agent to collect that piece of gold [\emph{in} \cite{hubner2008developing,bordini2006bdi}].''
The default \JASON\ intention selection me\-thod, round-robin based, is not much context aware. While extensions like plans ``priority'' annotations can provide cues to more informed choices, action selection is the subject of an huge area of probabilistic \ac{AI} centred around the \ac{MEU} principle. Within the \ac{PGM} setting this principle is instantiated by influence diagrams, graphical models extended with special nodes to represent utilities and actions.
\subsection{\uppercase{Resolution Path}}
%
%	Start this (sub)section with a nice figure.
%
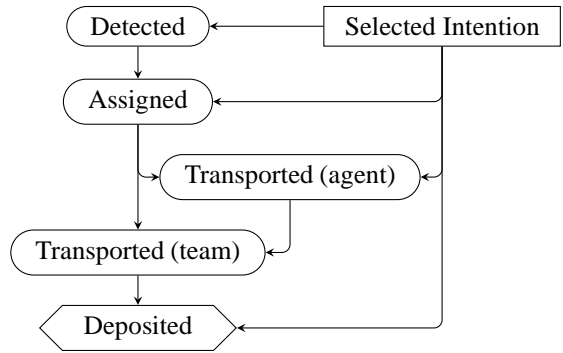
\begin{figure}[t]
\begin{center}
\begin{tikzpicture}
\node[utilnode] at (0,0) (DG) {Deposited};
\node[varnode] at (0,1) (TTG) {Transported (team)};
\node[varnode] at (2,2) (ATG) {Transported (agent)};
\node[varnode] at (0,3) (SG) {Assigned};
\node[varnode] at (0,4) (FG) {Detected};
%\node[varnode, fill=gray!50!white] at (-2,3) (HG) {Hidden };
\node[actionnode] at (4,4) (A) {Selected Intention};

\path[->, every edge/.style={cond}]
%	(SG) edge (ATG)
%	(ATG) edge (TTG)
	(SG) edge (TTG)
	(TTG) edge (DG)
	(FG) edge (SG)
	(A) edge (FG)
%	(A) edge (SG)
%	(A) edge (ATG)
%	(A) edge[bend left] (DG)
;
\draw[cond,->,rounded corners=3pt] (A) |- (SG);
\draw[cond,->,rounded corners=3pt] (A) |- (ATG);
\draw[cond,->,rounded corners=3pt] (SG) |- (ATG);
\draw[cond,->,rounded corners=3pt] (ATG) |- (TTG);
\draw[cond,->,rounded corners=3pt] (A) |- (DG);
\end{tikzpicture}
\end{center}
\caption{%
An influence diagram for intuition selection of \GM\ agents. The discovery, auctioning, transportation and deposit of gold pieces of the coordination protocol is represented in the graph. Variables (denoted by round rectangles) and Utilities (chamfered rectangles) in this diagram refer to \emph{total quantities} of gold pieces. The probabilistic effect of the selected action is represented by the arrows that leave the action (rectangular) node.
}
\label{fig:influence.diagram.1}
\end{figure}
\noindent
Given the \GM\ scenario the miners intention selection function is described using an influence diagram as depicted in figure \ref{fig:influence.diagram.1}. Deposited golds define an utility node and the range of actions is extracted form active intentions. The miners team coordination protocol is represented from the start with the discovery of a new gold piece to termination with the deposit of that gold piece.

The resulting utility function can then be used by the MEU principle to select, from the available action (the heads of instantiated plans in the intentions stack), the optimal one. Nodes in the influence diagram represent \acp{CPD} that, as part of the resolution proposed here, are to be tuned. Once the influence diagram defined, existing java \ac{PGM} libraries that supports influence diagrams (e.g. \SAMIAM) generate a (static) deliberation policy that can be inserted into the agent’s intention selection function defined in \JASON.

\section{\uppercase{Conclusion}}
\noindent
Looking forward the \GM\ exercise two issues require further consideration:
\begin{itemize}
\item
``real-time'' and ``off-line'' symbolic/probabilistic levels interplay;

\item
utility/goal conflicts;

\item
multi-agent applications;
\end{itemize}

In the deliberation process the communication between symbolic and probabilistic levels is of key importance. Selection functions signatures already define some ``real-time'' channels: arguments carry information from the symbolic to the probabilistic level while the returned values work the other way around. But perhaps other forms of mutual ``off-line'' influence can be considered. For example, a process where at the probabilistic level influence diagrams (e.g. \ac{BN}) are structured from the set of symbolic plans and beliefs and the coefficients in the factors of the nodes are learned from experience. One can also imagine a similar influence going from the probabilistic to the symbolic level, either introducing relevant concepts from unsupervised feature learning \cite{socher2011parsing} using \ac{DNN} \cite{hinton2007learning,salakhutdinov2012efficient} or summarizing \ac{ML} techniques like \ac{PCA} \cite{jolliffe2005principal}.

The two-layer approach also raises some practical and theoretical concerns on behaviour consistency. Since the probabilistic level maximizes utility functions and the symbolic level derives intentions from goals one as to ask how the utilities relate with the goals: if they are inconsistent the agent will be in trouble — like HAL 9000, as explained in the movie ``2010''.

Multi-agent systems imply social communication that in \JASON\ is represented by messages that must be selected from an in-box and composed and sent to the out-box. Here the probabilistic layer can contribute to the social perception in other ways besides the computation of the select message function. For example, by using \acp{BN} to represent reputation and expectations about other agents then bayesian learning updates the social reasoning.

Coupling symbolic and probabilistic AI is an hard task. Our approach tries to minimize and loosely regulate contact by the separation of competences. Such separation seems easier in a structured framework as \JASON\ than in a (somewhat) simpler but much broader setting as \ac{PGM} or \ac{PLP}. The proposed approach seems technically feasible and with the potential to lead to a synergy in both flavours of AI.

\subsection*{Future work}

\begin{itemize}
	\item Apply the two-layer approach outlined here in a virtual scenario (\emph{eg} \GM);
	\item Consider probabilistic correction on other elements of the \ac{BDI} process. In particular, as a pre-process of the Belief-Revision function;
	\item Formulate a probabilistic version \ac{BDI} using \ac{SRL} or \ac{PLP};
\end{itemize}

\section*{\uppercase{Acknowledgements}}
\noindent
The people around us, the flow of experiences, the internet.

\bibliographystyle{ieeetr}
\bibliography{dapaiper}

\end{document}